\documentclass[PRD,reprint,twocolumn,showpacs,
amssymb, amsmath, aps, showpacs, nofootinbib, superscriptaddress]{revtex4}

\usepackage{multirow}
\usepackage{hyperref}
\usepackage{graphicx}
\usepackage{dcolumn}
\usepackage{bm}
\usepackage{threeparttable}
\usepackage{subfig}
\usepackage{color,txfonts}

\def\beq{\begin{equation}}
\def\eeq{\end{equation}}
\def\bey{\begin{eqnarray}}
\def\eey{\end{eqnarray}}

\def\lsim{\mathrel{\raise.3ex\hbox{$<$\kern-.75em\lower1ex\hbox{$\sim$}}}}
\def\gsim{\mathrel{\raise.3ex\hbox{$>$\kern-.75em\lower1ex\hbox{$\sim$}}}}

\begin{document}

\title{Neutron star-black hole coalescence rate inferred from macronova/kilonova observations}
\affiliation{Key Laboratory of Dark Matter and Space Astronomy, Purple Mountain Observatory, Chinese Academy of Sciences, Nanjing 210008, China}
\affiliation{Max Planck Institute for Gravitational Physics, Albert Einstein Institute, Callinstra\ss e 38, D-30167 Hannover, Germany.}
\affiliation{TianQin Research Center, Sun Yat-sen University, Zhuhai, 519082, China}
\affiliation{Tsinghua University, 30 Shuangqing Rd, Beijing, 100084, China.}
  \author{Xiang Li }
\affiliation{Key Laboratory of Dark Matter and Space Astronomy, Purple Mountain Observatory, Chinese Academy of Sciences, Nanjing 210008, China}
\author{Yi-Ming Hu}
\email{yiming.hu@aei.mpg.de}%
\affiliation{Max Planck Institute for Gravitational Physics, Albert Einstein Institute, Callinstra\ss e 38, D-30167 Hannover, Germany.}
\affiliation{TianQin Research Center, Sun Yat-sen University, Zhuhai, 519082, China}
\affiliation{Tsinghua University, 30 Shuangqing Rd, Beijing, 100084, China.}
\email{yiming.hu@aei.mpg.de (YMH)}%
  \author{Zhi-Ping Jin }
\affiliation{Key Laboratory of Dark Matter and Space Astronomy, Purple Mountain Observatory, Chinese Academy of Sciences, Nanjing 210008, China}
\author{Yi-Zhong Fan}
\email{yzfan@pmo.ac.cn}%
\author{Da-Ming Wei }
\affiliation{Key Laboratory of Dark Matter and Space Astronomy, Purple Mountain Observatory, Chinese Academy of Sciences, Nanjing 210008, China}


\begin{abstract}
  Neutron star$-$black hole (NS$-$BH) coalescences are widely believed to be promising gravitational wave sources in the era of advanced detectors of LIGO/Virgo but such binaries have never been directly detected yet. Evidence for NS$-$BH coalescences have been suggested in short and hybrid GRB observations, which are examined critically.
  Based on the suggested connection between the observed macronovae/kilonovae events and NS$-$BH coalescences, we get a fiducial lower limit of NS$-$BH coalescence rate density ${\cal R}_{\rm nsbh} \approx 18.8^{+12.5}_{-8.6} ~{\rm Gpc^{-3}~ yr^{-1}~ (\theta_j/0.1~{\rm rad})^{-2}}$, where $\theta_{\rm j}$ is the typical half-opening angle of the GRB ejecta.
  The real value of  ${\cal R}_{\rm nsbh}$ is likely at least $\sim {\rm a~few}$ times larger, depending upon the equation of state of NS material and the properties of the NS$-$BH system, such as the mass and spin distribution of the black hole.
  If the link between macronovae/kilonovae and NS$-$BH coalescence is valid, one can expect that at design sensitivity the aLIGO/AdVirgo network will detect NS$-$BH coalescence signals at a rate of at least a dozen per year, and to consequently place constraints on certain physical properties of NS$-$BH systems.
\end{abstract}

\pacs{04.30-w, 04.30.Db, 98.70.Rz}

\maketitle
\section {Introduction}
Compact binaries consist of neutron stars and/or black holes are widely believed to be promising sources of gravitational waves \cite{Clark1977}.
So far, ten binary neutron star (BNS) systems have been observed in the Galaxy, and two confirmed binary black hole (BBH) events with one extra possible detection have been directly detected in the first advanced LIGO observational run (O1), while no NS$-$BH binary has been directly observed yet \cite{Lattimer2012,Abbott2016,Abbott2016b,Abbott2016c}, thus the estimates of NS$-$BH coalescence rate can only be made indirectly.
For example an estimated rate density of ${\cal R}_{\rm nsbh}\sim 0.04-10^{3}~{\rm Gpc^{-3}~yr^{-1}}$ can be derived from stellar evolution synthesis \cite{Shibata2011,Narayan1991,Abadie2010,Dominik2014}.
Upper limits ($\sim 10^{3}~{\rm Gpc^{-3}~yr^{-1}}$) were also given from GRB observation assuming all GRBs are linked with NS$-$BH systems \cite{Nakar2007} and from the assumption that all the {\emph r}-process material were produced in NS$-$BH coalescences \cite{Bauswein2014}.

In addition to generating strong GW signal, a fraction of NS$-$BH coalescences are also expected to produce high energy transients, in particular supernova-less GRBs \cite{Narayan1992}, including short GRBs (sGRBs) and hybrid GRBs (hGRBs).
Therefore it is feasible to infer the NS$-$BH coalescence rate with the sGRB/hGRB observations.
The main challenge is however that both BNS and NS$-$BH coalescences could power short/hybrid GRBs and in almost all the stellar evolution synthesis-based estimates, the BNS coalescence rate is significantly higher than ${\cal R}_{\rm nsbh}$ \cite[e.g.][]{Abadie2010,Dominik2014}.
Moreover, the tidal disruption of the NSs in the NS$-$BH coalescences is necessary for generating electromagnetic (EM) transient emissions like GRBs, otherwise the NSs would have been wholly swallowed by the BHs. The probability of tidal disruption depends on the mass ratio ($\eta$) between the BH and NS, the equation of state (EOS) of the NS material, the initial dimensionless spin ($\chi$) of the BH, and the initial tilt angle of the binary system \cite{Shibata2011,Belczynski2008,Pannarale2014}.
All of these factors make the estimate of NS$-$BH merger rate from direct observations of sGRB more uncertain.
Thus, a more realistic estimate of NS$-$BH merger rate from observation relies on further physical uniqueness of NS$-$BH systems.

Some evidences for the NS$-$BH coalescence-driven short/hybrid GRBs have been suggested and we examine them {\it critically}: \\
(i) {\it The offset distribution argument.} With a statistical investigation of the spatial offsets of short/hybrid GRBs from their host galaxies, Troja et al. \cite{Troja2008} found an intriguing trend that the events with extended soft radiation components lasting for $\sim 100$ s lie very close to their
host galaxies. Such an extended-duration/low-offset group, consisting of $\sim 1/3$ of the sample investigated there, has been suggested to be caused by NS$-$BH coalescences \cite{Troja2008}. If correct, this would suggest a
${\cal R}_{\rm nsbh}\sim {\cal R}_{\rm merg,grb}/3f_{\rm grb} \sim 600~(2.5f_{\rm grb})^{-1}~{\rm Gpc^{-3}~yr^{-1}}$, where we have taken ${\cal R}_{\rm merg,grb} \sim 700~{\rm Gpc^{-3}~yr^{-1}}$ \cite{Fong2014} and the GRB production fraction $f_{\rm grb}\sim 0.4$ \cite{Belczynski2008}.
However, the studies with the Hubble Space Telescope sample do not find significant difference in the two offset distributions, and there seems no clear evidence from their
locations that sGRBs with and without extended emission
require different progenitor systems \cite{Fong2010,Fong2013a}.

(ii) {\it The under-luminous nearby sGRB ``excess" argument.} Recently  Siellez et al. \cite{Siellez2016} argued that six nearby (the redshift $z<0.3$) under-luminous sGRBs
were produced dynamically in globular clusters as a result of NS$-$BH mergers. If correct, a high ${\cal R}_{\rm nsbh}$  is favored \cite{Siellez2016}.
One {\it caution} is that the redshifts of four events in their sample are not as ``secure" as others studied in the literature (For GRB 070923 and GRB 090417A the redshifts are usually taken as unknown \cite{Berger2014}. The association of GRB 060502B and GRB 061201 with their ``suggested host galaxies" are just at the significance level of $\sim 2\sigma$ \cite{Fong2010,Fong2013a}).
Supposing the host galaxies taken in \cite{Siellez2016} are correct, the projected offset of  GRBs to their hosts are $\sim (64~{\rm kpc},~70~{\rm kpc},~34~{\rm kpc},~530~{\rm kpc},~18~{\rm kpc},~320~{\rm kpc})$ for (GRB 050509B, GRB 060502B, GRB 061201, GRB 070923, GRB 080905A, GRB 090417A) respectively \cite{Troja2008,Fox2007,Bloom2009,Fong2013a},
remarkably larger than the averaged values for short events \cite{Fong2010} and at odds with that suggested in \cite{Troja2008}.
Moreover, the candidates in \cite{Siellez2016} are short of optical/radio afterglow emission to reliably constrain the half-opening angle $\theta_{\rm j}$,
for which the rate estimate subjects to huge uncertainty.

(iii) {\it The large {\emph r}-process material mass argument.}
A macronova/kilonova could arise from the radioactive decay of heavy elements \cite{LiLX1998} produced in the coalescence of BNSs or NS$-$BH binaries \cite{Lattimer1974,Eichler1989}.
The major differences of the ejecta from these two type of progenitors are \cite{Hotokezaka2013}:
a) the NS$-$BH coalescences could eject much more material than the BNS coalescences and in current numerical simulations the former could eject material up to $\sim 0.2~M_\odot$ while the latter usually can only eject dynamical material with a mass $\lesssim 0.02~M_\odot$ \cite{Hotokezaka2013b,Kyutoku2015}; b) the NS$-$BH coalescence ejecta is concentrated along the disk place while the BNS coalescence ejecta is largely isotropic \cite{Tanaka2014}. Consequently the macronovae powered by some NS$-$BH coalescences can be much more luminous and bluer and last longer than those from BNS coalescences \cite{Tanaka2014}. A macronova model light curve generated from numerical simulation for the ejecta from an NS$-$BH coalescence, with a velocity $\sim 0.2c$ and mass $M_{\rm ej}\sim0.1M_{\odot}$, can reasonably reproduce the ``excess" displayed in hGRB 060614 \cite{Yang2015}. The same model works also for the macronova signal displaying in sGRB 050709 with a $M_{\rm ej}\sim 0.05~M_\odot$ \cite{Jin2016}.
To reproduce the macronova signal in sGRB 130603B within the BNS coalescence scenario, an unusual large $M_{\rm ej}\sim 0.03-0.08~M_\odot$ is needed \cite{Tanvir2013}.
Instead, such a signal can be naturally reproduced within the NS$-$BH coalescence scenario for a $M_{\rm ej}\sim 0.05~M_\odot$ \cite{Hotokezaka2013,Kawaguchi2016,Jin2016}.
Encouragingly, all three GRB/macronova events have very small offsets from their hosts, as suggested by Troja et al. \cite{Troja2008} for NS$-$BH coalescence events.

\section {Method}
Below we focus on the derivation of ${\cal R}_{\rm nsbh}$ from the GRBs displaying macronova signals. GRB 050709, GRB 060614 and GRB 130603B were observed at redshifts of $z=(0.16,~0.125,~0.356)$, respectively. Jet breaks had been reliably measured in GRB 060614 and GRB 130603B and $\theta_{\rm j}\approx 0.1~{\rm rad}$ were inferred \cite{Xu2009,Fong2014}. For GRB 050709 the recent analysis suggests that an early jet break appeared at $t\leq 1.4$ days \cite{Jin2016}, implying a $\theta_{\rm j}\leq 0.1~{\rm rad}$.

GRB 050709 was detected by HETE-II with a field of view ${\rm F.o.V}\approx 3\,{\rm sr}$  \cite{Villasenor2005} while GRB 060614 and GRB 130603B were recorded by {\it Swift} satellite with a ${\rm F.o.V}\approx 2.4\,{\rm sr}$  \cite{Gehrels2006,Tanvir2013}. Note that HETE-II has a much lower detection rate of sGRBs in comparison with {\it Swift} due to the relatively small effective area of the onboard detector. Moreover, no GRBs were reported by HETE-II any longer since March 2006 \footnote{href{http://space.mit.edu/HETE/Bursts/}}. For simplicity we ignore the difference between HETE-II and {\it Swift}.
The GRB-less macronovae were ignored when we talk about the rate of macronovae as all events identified so far were triggered from observation of GRB. The {\it rate density} of NS$-$BH driven macronovae can be calculated via
\begin{equation}
  {\cal R}_{\rm nsbh-mn}={{\cal N_{\rm nsbh-mn}} \over V_{\rm com}(z\leq 0.4)~\cal T}{4\pi \over {\rm F.o.V}}{1\over 1-\cos\theta_{\rm j}},
\end{equation}
where ${\cal N}_{\rm nsbh-mn}$ is the total number of detected macronovae that are believed to be linked to NS$-$BH coalescence, ${\cal T}$ is the observation time (in our case we take it as $\approx 11$ years since the sGRBs were firstly localised in 2005) and $V_{\rm com}(z\leq 0.4)$ is the comoving volume. In this work we consider $z\leq 0.4$ since at higher $z$ the detection of macronovae becomes very hard except for Hubble Space Telescope, while such followup observations of high-$z$ sGRBs were extremely rare.
Hence we have
\begin{eqnarray}
  {\cal R}_{\rm nsbh}&=&{{\cal R}_{\rm nsbh-mn}\over {\cal F}_{\rm nsbh-mn,z}f_{\rm nsbh-mn}}\approx 18.8^{+12.5}_{-8.6}~{\rm Gpc^{-3}~yr^{-1}}\left({\theta_{\rm j}\over 0.1~{\rm rad}}\right)^{-2}\nonumber\\
&&~\times{{\cal N_{\rm nsbh-mn}}\over 3}{\cal F}_{\rm nsbh-mn,z}^{-1}f_{\rm nsbh-mn}^{-1}\left({{\cal T} \over 11~{\rm yr}}\right)^{-1}\left({\rm F.o.V \over 2.4~{\rm sr}}\right)^{-1}
\end{eqnarray}
where $f_{\rm nsbh-mn}$ is the fraction of NS$-$BH mergers that can produce the observed macronova signal, in other words, it can launch un-bound material as heavy as $M_{\rm ej}\geq 0.05~M_\odot$, which is the smallest amount of {\emph r}-process material found in the modelling of the current 3 ``bright" macronovae.
The value of $f_{\rm nsbh-mn}$ is less than one, and the details of calculation is explained later.
${\cal F}_{\rm nsbh-mn,z}\leq 1$ is the detection ratio of the NS$-$BH coalescence-powered macronovae as a function of redshift.
As a {\it conservative} estimate on ${\cal R}_{\rm nsbh}$ below we take ${\cal F}_{\rm nsbh-mn,z}=1$.

For the Advanced LIGO detectors that can detect the gravitational wave radiation from NS$-$BH mergers within a typical distance $D\sim 400$ Mpc (for $1.4-10 M_\odot$ NS$-$BH system) in their full performance, the detection {\it rate} is roughly expected to be
\begin{equation}
  {R}_{\rm GW,nsbh}\approx 5.0^{+3.3}_{-2.3}~f_{\rm nsbh-mn}^{-1}\left({\theta_{\rm j}\over 0.1~{\rm rad}}\right)^{-2}\left({D\over 400~{\rm Mpc}}\right)^{3}~{\rm yr^{-1}}.
\label{eq:R_GWnsbh}
\end{equation}
Below we focus on the estimate of $f_{\rm nsbh-mn}$.
Similar quantities discussed in the literature are $f_{\rm grb}$ ($\sim 0.4$, \cite{Belczynski2008}) or the EM counterpart production fraction $f_{\rm em}$ ($\sim 1/3$, \cite{Pannarale2014}).
The difference is that the key parameter needed in \cite{Belczynski2008,Pannarale2014} is the accretion disk mass ($M_{\rm disk}$ \cite{Foucart2012}), while in our case it is $M_{\rm ej}$.
If we assume $f_{\rm nsbh-mn} \approx f_{\rm em}\sim 1/3$, then eq.(\ref{eq:R_GWnsbh}) reads ${R}_{\rm GW,nsbh} \sim 15.0^{+9.9}_{-6.9}~{\rm yr^{-1}}$. This estimate can be taken as a fiducial conservative reference which implies a promising detection prospect.

In order to calculate $f_{\rm nsbh-mn}$ we adopt an empirical formula for $M_{\rm ej}$ of NS$-$BH coalescence \cite{Kawaguchi2016}.
Moreover, considering that the black hole mass ($M_{\rm BH}$) and spin $\chi_{\rm BH}$ play an important role in generating electromagnetic counterparts and the detectable gravitational wave signals, eq.(\ref{eq:R_GWnsbh}) can be re-written into a general form,
\begin{equation}
R_{\rm GW,nsbh}={\cal R}_{\rm nsbh}\langle  V \rangle={\cal R}_{\rm nsbh-mn} {\langle  V \rangle \over  \langle P_{\rm mn} \rangle},
\label{eq:newRate-1}
\end{equation}
where $\langle  V \rangle =\int {\rm d}M_{\rm BH} P(M_{\rm BH})  \int {\rm d} \chi_{\rm BH} P(\chi_{BH}) V(M_{\rm BH},\chi_{\rm BH})$ and $\langle P_{\rm mn} \rangle=\int {\rm d}M_{\rm BH} P(M_{\rm BH}) \int {\rm d} \chi_{\rm BH} P(\chi_{BH}) P_{\rm mn}(M_{\rm BH},\chi_{\rm BH})$,
$\langle \cdot \rangle$ represents the weighted average over black hole mass $M_{BH}$ and black hole spin $\chi$, under certain mass distribution $P(M_{\rm BH})$ and spin distribution $P(\chi_{\rm BH})$,
$V$ is the sensitive volume of the advanced GW detector network. $P_{\rm mn}(M_{\rm BH})$ is the probability of a NS$-$BH coalescence produce a macronova signal with a given $M_{\rm ej}$. Here this probability is simply a Heaviside function of the {\emph r}-process material ejecta mass over a pre-determined threshold of $0.05M_\odot$.
Note that $\langle P_{\rm mn} \rangle=f_{\rm nsbh-mn}$.

We estimate $M_{\rm ej}$ through equations (1-4) from \cite{Kawaguchi2016}, and further deduce $P_{\rm mn}(M_{\rm BH},\chi_{\rm BH})$.
For simplicity, the neutron star's mass is fixed to $1.35~M_{\odot}$.
We discuss a variety of combinations of NS$-$BH systems' properties, including the equation of state (EoS) of the neutron star, the distribution of black hole's mass as well as spin. We demonstrate how change of properties of NS$-$BH systems might modify the predicted detection rate.

The EoS of neutron stars is very uncertain \cite{Oerte2016}. Following \cite{Kawaguchi2016}, we chose four EoSs, namely
{\it APR4, ALF2, H4} and {\it MS1}.
With a mass of $1.35 M_\odot$, the anticipated radii are $ 11.1,~12.4,~13.6,~14.4\, ~{\rm km}$ respectively.
We choose two ends of possible astronomical distributions as suggested by \cite{Abbott2016b}, namely the mass distributions of the black hole follows one of the following distributions:
(1) $P\big(\log(M_{\rm BH})\big) = {\rm constant}$;
(2) $P(M_{\rm BH})\propto M_{\rm BH}^{-2.35}$.
The minimum boundary could either be $3 M_\odot$ as motivated by the observed maximum mass of neutron star, or $5 M_\odot$ as is the observed minimum mass of black holes. These two masses does not equal, leading to the ``mass gap".
The maximum mass is chosen to be $99 M_\odot$ as this is the upper limit for LIGO detection pipelines.
Although no EM observation provided evidence for black holes heavier than $\sim 16 M_\odot$
gravitational wave observation showed clearly that stellar mass black hole as heavy as $\sim 60 M_\odot$ can exist, which endorses our choice.
As for the spin, we examine a ``flat" spin distribution $P(\chi) = U[-1,1]$ (labeled spin distribution $1$), and compare against a ``bimodal" distribution \cite{Stone2013} $P(|\chi|) = U([0,0.3]\cup[0.7,1])$ (labeled spin distribution $2$) hinted by current observations of stellar mass black hole spin.
For ``high spin" case where one assume all black holes has an absolute value of spin $|\chi|> 0.7$, the estimated rate would be reduced by a factor of $\sim 2$ from the bimodal case.

In this study, we assume an (anti-)aligned spin for the black hole, so misalignment is not discussed.
Otherwise we need to introduce an extra level of assumptions, which lacks solid astronomical prior.
Notice that the population synthesis of NS$-$BH binaries suggested that most of the systems have a relatively small spin-orbit misalignment \cite{Belczynski2008}.

\section {Results}
Based on the above combinations, we can predict detection ability of two aLIGO detectors as well as the coalescence rate of the population of NS$-$BH systems.
Notice that all the quantitative conclusions rely on several assumptions.
We assume all observed macronovae have an NS$-$BH origin, a minimum ejected mass of $0.05M_\odot$ and complete macronovae observations under redshift $0.4$.
We discuss the influence of invalidity of such assumptions later.

\begin{table*}
  \small
  \centering
  \caption{Expected detection rate (in $\times10^2 \, \rm yr^{-1}$) as well as astronomical rate density of NS$-$BH systems (in $\rm \times 10^2 \,Gpc^{-3}~ yr^{-1}$) in the advanced GW detector era. Different combinations of EoS, minimum black hole mass, mass distribution and spin distribution would lead to different rates.}\label {tab:rates}
  \subfloat[detection rates in $\rm \times 10^2\,yr^{-1}$]{
    \hspace{.5cm}
\begin{tabular}{|c|c|c|cccc|}
  \hline
  $M^{\rm min}_{\rm BH}$    & distribution   & $\chi_{\rm BH}$  & APR4       &   ALF2    &   H4    &   MS1 \\ \hline
  \multirow{4}{*}{$3M_\odot$}  & \multirow{2}{*}{1}
   &flat            & $22.6^{+14.9}_{-10.3}$  & $3.2^{+2.1}_{-1.5}$ &$2.2^{+1.4}_{-1.0}$&  $1.3^{+0.9}_{-0.6}$\\
   & &bimodial      & $13.8^{+9.1}_{-6.3}$  &  $2.0^{+1.3}_{-0.9}$ & $1.3^{+0.9}_{-0.6}$& $0.8^{+0.5}_{-0.4}$\\
   \cline{2-7}
   &\multirow{2}{*}{2}
       &flat         & $10.3^{+6.8}_{-4.7}$ & $0.8^{+0.6}_{-0.4}$ & $0.6^{+0.4}_{-0.3}$ & $0.3^{+0.2}_{-0.1}$\\
   &    &bimodial & $6.3^{+4.1}_{-2.9}$ & $0.5^{+0.3}_{-0.2}$  &$0.4^{+0.2}_{-0.2}$ & $0.2^{+0.1}_{-0.1}$\\\hline
  \multirow{4}{*}{$5M_\odot$}  & \multirow{2}{*}{1}
       &flat  & $21.6^{+14.3}_{-9.9}$& $4.0^{+2.6}_{-1.8}$ & $2.7^{+1.8}_{-1.3}$ & $1.8^{+1.2}_{-0.8}$\\
   &    &bimodial & $13.2^{+8.7}_{-6.0}$&  $2.4^{+1.6}_{-1.1}$ & $1.7^{+1.1}_{-0.8}$ &$1.1^{+0.7}_{-0.5}$\\
   \cline{2-7}
   & \multirow{2}{*}{2}
       &flat      & $7.5^{+5.0}_{-3.4}$ &  $1.1^{0.7}_{-0.5}$ & $0.8^{+0.5}_{-0.4}$ &$0.5^{+0.3}_{-0.2}$\\
   &    &bimodial & $4.6^{+3.0}_{-2.1}$ &  $0.7^{+0.4}_{-0.3}$ & $0.5^{+0.3}_{-0.2}$  & $0.3^{+0.2}_{-0.1}$\\\hline
\end{tabular}
  }\subfloat[rate density in $\rm \times 10^2\,Gpc^{-3}~ yr^{-1}$]{
    \hspace{.5cm}
\begin{tabular}{|c|c|c|cccc|}
  \hline
  $M^{\rm min}_{\rm BH}$    & distribution   & $\chi_{\rm BH}$  & APR4       &   ALF2    &   H4    &   MS1 \\ \hline
  \multirow{4}{*}{$3M_\odot$}  & \multirow{2}{*}{1}
  &flat                  & $64.4^{+42.5}_{-29.4}$   &$9.1^{+6.0}_{-4.2}$ &$6.2^{+4.1}_{-2.4}$& $3.7^{+2.5}_{-1.7}$\\
   & &bimodial & $38.6^{+25.5}_{-17.7}$  & $5.5^{+3.6}_{-2.5}$ & $3.7^{+2.5}_{-1.7}$ & $2.2^{+1.5}_{-1.0}$\\
   \cline{2-7}
   &\multirow{2}{*}{2}
       &flat  & $65.6^{+43.3}_{-30.0}$ & $5.4^{+3.6}_{-2.5}$ & $3.7^{+2.5}_{-1.7}$ & $2.0^{+1.3}_{-0.9}$\\
   &    &bimodial &$39.4^{+26.0}_{-18.0}$  & $3.3^{+2.2}_{-1.5}$ & $2.2^{+1.5}_{-1.0}$ & $1.2^{+0.8}_{-0.5}$\\\hline
  \multirow{4}{*}{$5M_\odot$}  & \multirow{2}{*}{1}
       &flat  & $54.6^{+36.0}_{-25.0}$ &$10.0^{+6.7}_{-4.6}$  & $6.9^{+4.6}_{-3.2}$& $4.6^{+3.0}_{-2.1}$\\
   &    &bimodial& $32.8^{+21.6}_{-15.0}$& $6.0^{+4.0}_{-2.8}$ & $4.2^{+2.8}_{-1.9}$ & $2.7^{+1.8}_{-1.3}$\\
   \cline{2-7}
   & \multirow{2}{*}{2}
       &flat & $32.6^{+21.5}_{-14.9}$&  $4.7^{+3.1}_{-2.2}$ &$3.3^{+2.2}_{-1.5}$ & $2.1^{+1.4}_{-1.0}$\\
   &    &bimodial & $19.6^{+12.9}_{-9.0}$ &$2.8^{+1.9}_{-1.3}$ & $2.0^{+1.3}_{-0.9}$ &$1.3^{+0.9}_{-0.6}$\\\hline
\end{tabular}
  }
\end{table*}

The detailed results is shown in table \ref{tab:rates}.
We can observe a large variety of rates with different parameter combinations.
However, overall, a rate of more than $\sim 20$ detections per year was expected for the two aLIGO detectors.
Even a duty cycle of $80\%$ for each detector can still guarantee a dozen of detections per year.
The joining of other detectors like AdV, KAGRA and LIGO-India would only increase this rate.

When all other conditions are equivalent, the expected detection rate of NS$-$BH systems depends strongly on the EoS of neutron star.
Between the most extreme cases, the rates could vary by a factor of $\sim 30$.
This is not surprising since neutron stars with a softer EoS is harder to eject enough matter outside the final black hole, thus a number of 3 detection of macronova implies a huge amount of NS$-$BH mergers; on the other hand, a stiffer EoS can support a higher fraction of macronova for NS$-$BH coalescence, thus decreasing the anticipated detection rate.

Notice that the uncertainty in our result is dominated by numerical fluctuation caused by a small number macronovae detections.
If future observations verify our assumptions, EM observations triggered EM follow-ups of NS$-$BH mergers can alleviates uncertainties in our estimation.
More GW detections can also measure BH mass and spin distribution.
By comparing calculated rates against GW detections rates, the EoS of the neutron star could be constrained.

Future EM facilities are expected to constrain neutron star EoS relatively well (\emph {e.g.}~\cite{Gendreau2012}.)
If we trust the EoS from EM observation, the perspective can be shifted towards independently verifying the distribution of black hole spin $\chi_{\rm BH}$ in a NS$-$BH system.
Higher spin of black hole implies a smaller innermost stable circular orbit (ISCO), thus more mass ejected.
All other properties being equivalent, NS$-$BH systems with different spin distributions have different rates.
By comparing actual detection rates against the calculated values, it is possible to put an independent constraints on the spin distribution, thus to either verify or disprove the bimodality of spin distribution.

The mass parameters can also be constrained, but not as well as other parameters.
Essentially the mass gap has negligible effect on the rate.
Although the apparent potential of distinguishing mass distributions is even better than for spin distribution, one should not simply take it at face value.
Firstly, the mass distribution is chosen to be enveloping real distribution, the actual distribution is less clear {\it a prior}.
Furthermore, the mass parameters can be relatively well constrained from GW data analysis, thus the mass distributions might be better determined by accumulating detections.
This is different from the spin scenario where the spin parameter might be only poorly constraint individually from GW data analysis.

\begin{figure}[!tbp]
    \centering
    \subfloat[advanced GW detector detection rate  ]{\includegraphics[width=0.5\textwidth]{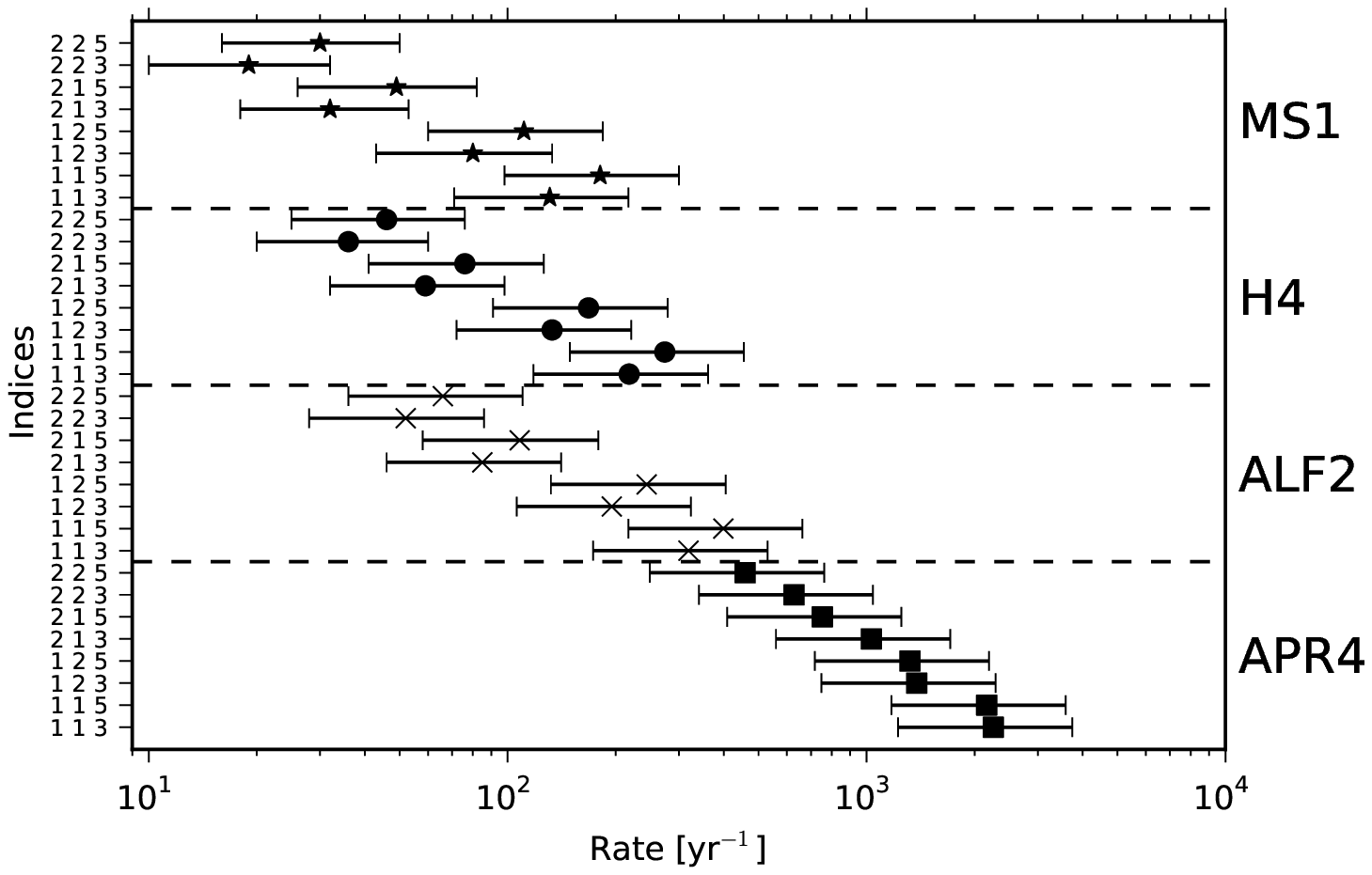}\label{fig:DetRate}}
    \hfill
    \subfloat[rate density and comparison with observation]{\includegraphics[width=0.5\textwidth]{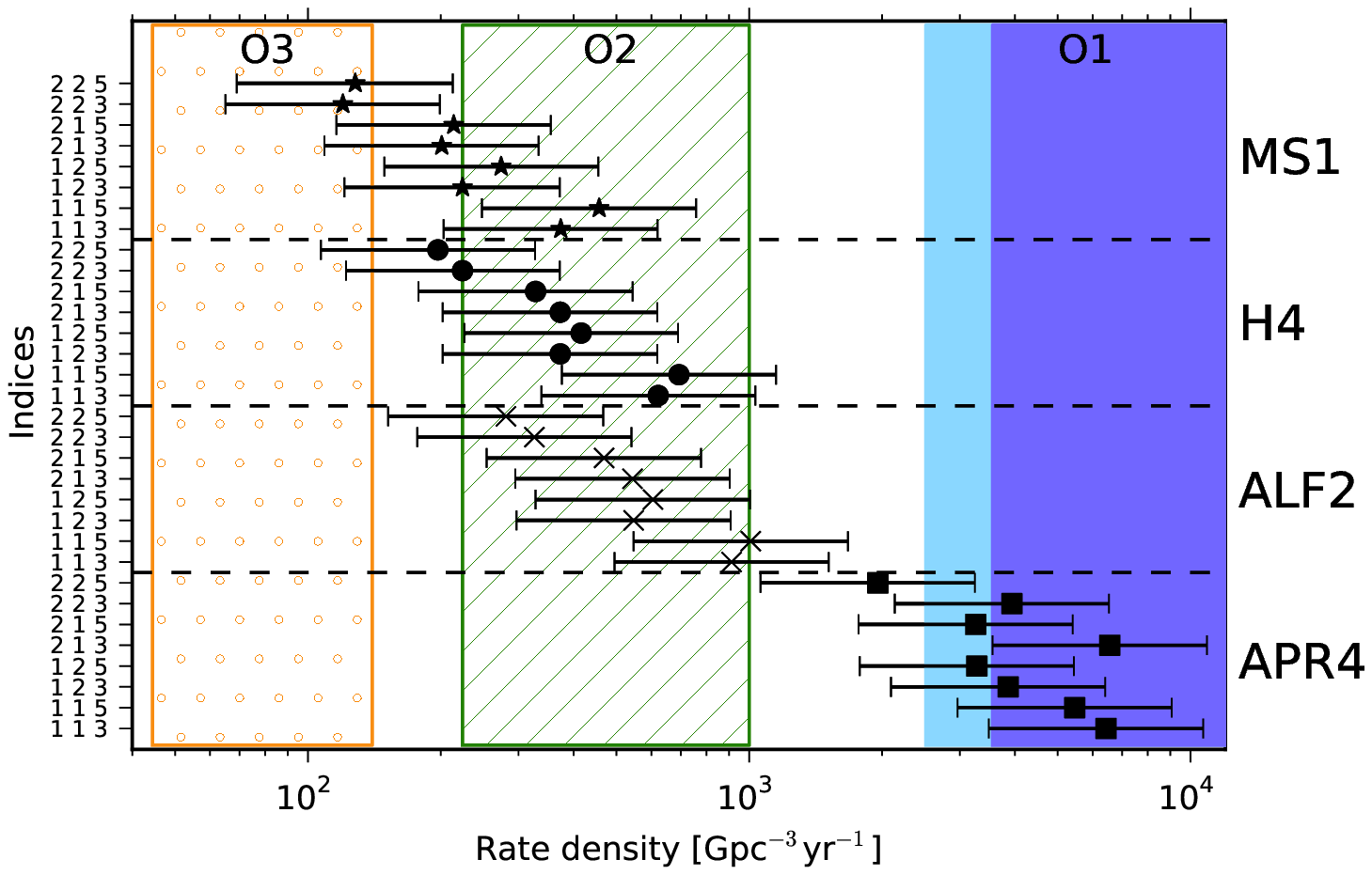}\label{fig:RateDens}}
  \caption{The numbers in the left label represents mass distribution index ($1$ for uniform over logarithm, $2$ for power law with power of $-2.35$), spin distribution index ($1$ for flat, $2$ for bimodal) and lower mass limit (in $M_\odot$) respectively. The three shaded regions represents a $90\%$ upper limit deduced (for O1) or can be deduced (for O2 and O3) from null observation of NS$-$BH. The lighter blue in O1 represents conclusion from $1.4-5 M_\odot$ while darker blue for $1.4-10 M_\odot$ NS$-$BH systems.\label{fig:rates}}
\end{figure}

We plot the expected detection rate $R_{\rm GW,nsbh}$ from advanced GW detectors in figure \ref{fig:DetRate}, the corresponding rate density ${\cal R}_{\rm nsbh}$ in figure \ref{fig:RateDens}.
In figure \ref{fig:RateDens} the upper limits of O1, O2 and O3 from \cite{Abbott2016c} were also plotted, assuming no detections in these periods.
In each plot, the estimated rates (density) together with uncertainties were plotted for different EoS, black hole mass and spin distributions.
Different parameters would lead to a large difference, which collectively covers two orders of magnitude.
However, for a given combination of parameters, the uncertainty is much narrower (around a factor of 3), which is also expected to be quickly shrinking with future observations.

\section {Conclusion}
To sum up, assuming NS$-$BH coalescences are responsible for the three bright macronovae/kilonovae, one can estimate the NS$-$BH merger rate.
This estimate is sensitive to physical parameters like EoS of neutron star and mass/spin distribution for black hole.
By comparing detection rate of NS$-$BH systems from aLIGO/AdV detectors and prediction from table \ref{tab:rates}, one can put constraints on the general properties of NS$-$BH systems.
With the non-observation of such systems in O1, an upper limit of $3,600~{\rm Gpc^{-3}~ yr^{-1}}$ is drawn assuming $1.4-5 M_\odot$ NS$-$BH system\cite{Abbott2016c}, which disfavours EoS APR4.
Other EoSs of the NS encounter no difficulty explaining the O1 non-detection.
Currently the uncertainty is too large to distinguish between different spin and mass features, however, such uncertainty is caused by the small number of macronovae detections.
Future GW observations and EM followups are promising in observing more macronovae, and would quickly decrease the associated error.

There are some caveats in this work.
For example we assume the completeness ${\cal F}_{\rm nsbh-mn,z}= 1$, which might not be correct by a factor of $2$.
All the conclusions we derived rely on the validity of the assumption that all observed macronovae were associated with NS$-$BH coalescences
(The coalescences of eccentric BNSs may be able to launch very-massive ejecta and yield bright macronovae but generate very different
 gravitational wave signals from that assume no eccentricity \cite{Baiotti2016}. 
 A lower limit of $\sim 20~{\rm Gpc^{-3}~ yr^{-1}}$ for eccentric BNS mergers can be concluded if all luminous macronovae were linked to such mechanism.)
%
However, as long as some of the macronovae are related with NS$-$BH coalescence, the general methodology can still be used to constrain NS$-$BH properties, under the condition that one can distinguish macronovae related with NS$-$BH mergers from BNS mergers.
The choices of some parameters can also affect the estimated rate, for example, if we change the mass threshold to $0.03 M_\odot$ the rate would be halved.
Also the current empirical equation on ejected mass from NS$-$BH was mainly calibrated on relatively small mass ratio (with $M_{\rm BH} < 10 M_\odot$), and with spin $\chi_{\rm BH}<0.9$.
So the extrapolation of mass estimation to a larger parameter space might not be guaranteed to be correct.
However, we only concern whether the pre-determined threshold is passed, the correctness of such decision should largely remain unaffected.
The consideration of spin-orbit misalignment might make the estimation more accurate, and we leave this issue for future consideration.

\acknowledgments  This work was supported in part by 973 Programme of China (No. 2013CB837000 and No. 2014CB845800), by NSFC under grants 11525313 (the National Natural Fund for Distinguished Young Scholars), 11273063 and 11433009, by the Chinese Academy of Sciences via the Strategic Priority Research Program (No. XDB09000000) and the External Cooperation Program of BIC (No. 114332KYSB20160007).


\end{document}